\documentclass[aps,prd,twocolumn,showpacs,nofootinbib]{revtex4-1}

\usepackage[latin1]{inputenc}
\usepackage{float}
\usepackage{amsmath,amssymb}
\usepackage{mathrsfs}
\usepackage{theorem}
\usepackage{graphicx}
\usepackage{color}
\usepackage{wasysym}
\usepackage{hyperref} 

\definecolor{blue}{rgb}{0,0,1}
\definecolor{green}{rgb}{0,1,0}
\definecolor{red}{rgb}{1,0,0}
\definecolor{vio}{rgb}{1,0,1}
\definecolor{ama}{rgb}{1,1,0}
\newcommand{\bc}{\begin{center}}
\newcommand{\ec}{\end{center}}
\newcommand{\be}{\nopagebreak[3]\begin{equation}}
\newcommand{\ee}{\end{equation}}
\newcommand{\ba}{\nopagebreak[3]\begin{eqnarray}}
\newcommand{\ea}{\end{eqnarray}}

{\theorembodyfont{\sffamily} }
{\theorembodyfont{\sffamily} \newtheorem{teo}{Th.}[section]}
{\theorembodyfont{\sffamily} }
{\theorembodyfont{\sffamily} }
{\theorembodyfont{\sffamily} }

\begin{document}

\title{\bf 
New derivation for the equations of motion for particles
in electromagnetism
}

\author{
Emanuel Gallo
and
Osvaldo M. Moreschi
}

\affiliation{FaMAF, Universidad Nacional de C\'{o}rdoba, \\
Instituto de F\'\i{}sica Enrique Gaviola (IFEG), CONICET, \\
Ciudad Universitaria, (5000) C\'ordoba, Argentina. }


\begin{abstract}
We present equations of motion for charged particles using balanced equations, and without 
introducing explicitly divergent quantities. This derivation contains as particular cases 
some well known equations of motion, as the Lorentz-Dirac equations. 

An study of our main equations in terms of order of the interaction with the external field
conduces us to the Landau-Lifshitz equations.
We find that the analysis in second order show a special behavior.
We give an explicit presentation up to  third order of our main equations,
and expressions for the calculation of general orders.

\end{abstract}


\pacs{03.30.+p, 03.50.De, 41.60.-m}

\maketitle


\section{Introduction}

The main question we would like to study in this article is how far can the notion
of a particle, as a point like object be extended, in the realm of classical electromagnetism.
Of course in the textbooks one encounters extensive discussions of test particles; which are normally used
for the very definition of the electromagnetic fields; namely, the electromagnetic fields
are those that enter into the Lorentz force acting on test particles.
However, the physical system becomes more complicated when one considers finite charged
particles. On the one hand, such a particle radiates, and therefore the equations
of motion should reflect the loss of energy-momentum of the particle.
On the other hand. a finite point like source involves fields that have a divergent
behavior as one approaches the particle; which in turn would imply a divergent
contribution to the total stress energy-momentum tensor of the system.

This problem has been tackled in the past using different techniques, 
as for example the idea of studying `spheres' in the limit when they become smaller and smaller.
In fact this guided the early works of Lorentz, Abraham, and 
Dirac\cite{Dirac38}.
In this article we present a discussion of the problem in which we avoid dealing
with infinite terms.
Our final result is a generalization of the equations found in the past, which contain
previous works as particular choices.

In the next section we present the basic notation used below.
Section \ref{sec:balance} is devoted to the presentation of the balance equations; which is
our main tool in this work.
The deduced equations of motion are a generalization of other similar equations found in the literature.
We also present in this section a generalized notion of total momentum for the charged particle.
Notably the new equations of motion involves new degrees of freedom, that are discussed below.
A particular choice of the new degrees of freedom conduces to the celebrated Lorentz-Dirac
equations of motion.
In section \ref{sec:generalsolutions} we study the nature of the solutions to the general problem.
Those solutions present difficulties in the physical interpretation, which motivate us to
study  the equations of motion in terms of orders of the strength of the interaction
with the external fields; which is presented in section \ref{sec:orders}.
In the last section we summarize our results.
An appendix is added where some properties of the solutions are presented.

\section{Dealing with particles in electromagnetism}
\subsection{Basic equations of a test charged particle}

Let us start by quickly reviewing the main equations of electromagnetism.
The dynamics of the electromagnetic fields is governed by Maxwell equations;
\begin{equation}
\label{eq:maxdivrel}
\nabla_{a} F^{ab}
=
\kappa J^b  ;
\end{equation}
and
\begin{equation}
\label{eq:fantisim0}
\nabla_{[a} F_{bc]} = 0 .
\end{equation}
While the dynamic of \emph{test particles} is determined by the Lorentz force
$f^a$ given by
\begin{equation}
\label{eq:fuerzaLrel}
f^a = q  F^{ab} v_b ;
\end{equation}
where one is considering a test particle of charge $q$ and four-velocity $v^a$.

The immediate question is: what is the force that acts on a particle? (not test particle)

The nature of the problem is that a (nontest) particle radiates if it is accelerated;
and therefore the use of the Lorentz force will imply an imbalance of energy
and momentum.
Several approaches have appeared to answer this question, and Lorentz, Abraham and 
Dirac\cite{Dirac38} have made important contributions to this goal, which have ended
in what is widely known as the Lorentz-Dirac equations of motion(see below).
We here present these well known equations of motion as a special case of our equations
by demanding balance of energy and momentum at an asymptotic sphere
defined by the intersection of the future null cone of the particle with future null infinity.

\subsection{Notation associated with a timelike curve}
Let us start with a massive point particle with mass
$m_A$ whose trajectory, in a flat space-time $(M,\eta_{ab})$,
is given by the timelike curve $C$,
which in a Cartesian coordinate system $x^a$ reads
\begin{equation}
\label{eq:trajectoria} x^a =z^a(\tau),
\end{equation}
with $\tau$ meaning the proper time of the particle along $C$.

The associated 4-velocity of this particle is
\begin{equation}\label{eq:4-velocity} 
v^a =\frac{dz^a}{d\tau},
\end{equation} 
and the signature of the flat metric is chosen such that $v_a v^a=1$.
(Note that we are using units in which $c=1$, so that $v^a$ has no units.)
Now, for each point $Q=z(\tau)$ of $C$, we draw a future null cone
$\mathscr{C}_Q$ with vertex in $Q$. If we call $x^a_P$
the Minkowskian coordinates of
an arbitrary
point on the cone $\mathscr{C}_Q$, then we can define
the retarded radial distance on the null cone by
\begin{equation}
\label{eq:retardedr} r = v_a \left(x^a_P-z^a(\tau) \right).
\end{equation}
A null vector
$l^a$ is defined by
\begin{equation}
\label{eq:nullvector} l^a=\frac{x^a_P-z^a(\tau)}{r}.
\end{equation}
And since $l^a$ is null, one has
\begin{equation}
 l_a \, l^a = 0 .
\end{equation}
Then one can see that
\begin{equation}
\label{eq:contrac1} v_a \, l^a = 1.
\end{equation}

One can introduce null polar coordinates associated to the timelike curve $C$ 
in the following way. 
Let $u$ be the null coordinate which is constant on the future null cones
defined for points $Q$ of $C$ and such that $u=\tau(Q)$.
Let $(\zeta,\bar\zeta)$ be stereographic coordinates of the sphere of directions
at the point $Q$.
Then, the relation between Minkowskian coordinates $x^a$
and null polar coordinates $(u,r,\zeta,\bar\zeta)$ is given by
\begin{equation}
 x^a = z^a(u) + r l^a(u,\zeta,\bar\zeta) .
\end{equation}

Let us note that by defining $\hat l^a$ to be the null vector
that corresponds to a stationary motion; which implies that 
$\hat l^a=\hat l^a(\zeta,\bar\zeta)$, one can see that 
$l^a$ and $\hat l^a$ will be proportional. In particular defining
\begin{equation}
 V(u,\zeta,\bar\zeta) = v^a(u) \hat l_a(\zeta,\bar\zeta) ;
\end{equation}
one can see that
\begin{equation}
 l_a(u,\zeta,\bar\zeta) = \frac{1}{V(u,\zeta,\bar\zeta)} \hat l_a(\zeta,\bar\zeta) .
\end{equation}

\subsection{Electromagnetic fields of a nontest charged particle}
The retarded electromagnetic field of a particle with charge $e$,
can be given in terms of the potential and/or the electromagnetic
field itself.

The retarded potential is
\begin{equation}
 A^a(x) = \frac{e v^a(u)}{r} .
\end{equation}

The corresponding electromagnetic field is\cite{Frolov79}
\begin{equation}
\begin{split}
 F_{ab} &= 2 e \left( 
\frac{1}{r V} \hat l_{[a} \dot V_{b]}
+
\frac{1}{r^2 V} ( 1 - \frac{r \dot V}{V}) \hat l_{[a}  V_{b]}
\right) \\
&= 2 e \left( 
\frac{1}{r} \left[ l_{[a} \dot V_{b]} - \frac{\dot V}{V} l_{[a}  V_{b]}\right] 
+
\frac{1}{r^2}  l_{[a}  V_{b]}
\right) 
;
\end{split}
\end{equation}
where a dot means derivative with respect to $u$ and we have chosen Gaussian units;
so that $\kappa = 4\pi$.

It is convenient to also have at hand the spinor components of the electromagnetic field;
which are given by
\begin{equation}
 \phi_0 = 0 ,
\end{equation}
\begin{equation}
 \phi_1 = -\frac{e}{2 r^2} ,
\end{equation}
and
\begin{equation}
 \phi_2 = -\frac{e V}{r} \bar\eth\left( \frac{\dot V}{V}\right) . 
\end{equation}
The symbol $\eth$ denotes the edth operator\cite{Geroch73} of the unit sphere,
and $\bar\eth$ is its complex conjugate.

\section{Balance equations for a non-test charged particle}\label{sec:balance}
\subsection{The conservation law for the total energy momentum tensor}

The main tool to derive the balanced equations of motion for charged particles
is the conservation law of the total energy-momentum tensor; namely
\begin{equation}\label{eq:conservT}
 \nabla \cdot T = 0 .
\end{equation}

Let $K_{\underline{c}}^b$ be four translational Killing vectors;
where we are using the numeric index $\underline{c}=0,1,2,3$ and the abstract index
$b$. Then from the conservation law (\ref{eq:conservT}) one also obtains that
\begin{equation}\label{eq:conservT2}
 \nabla_a \left( T^a_{\;\;b} \; K_{\underline{c}}^b \right) = 0 .
\end{equation}

\begin{figure}[htbp]
\centering
\includegraphics[clip,width=0.5\textwidth]{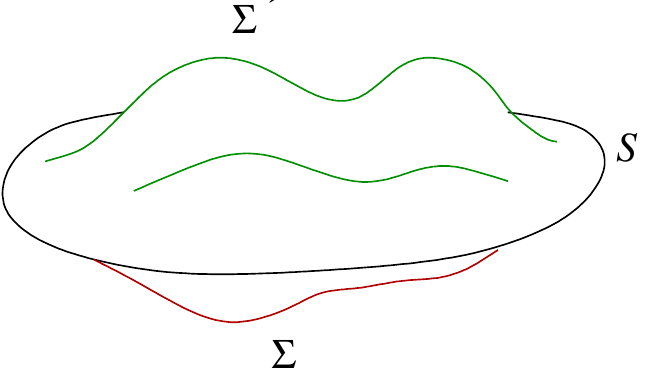}
\caption{Two spacelike hypersurfaces, $\Sigma'$ and $\Sigma$, are the boundary of the internal four volume 
$\mathscr{V}$.
In turn, each of these spacelike hypersurfaces has the same two dimensional boundary $S$.
}
\label{fig:sigma-sigmap}
\end{figure}

Then, let $\mathscr{V}$ be the four volume which has as boundaries the two spacelike hypersurfaces,
$\Sigma'$ at its future boundary and $\Sigma$ at its past boundary, as depicted in 
figure \ref{fig:sigma-sigmap}.
Also, each spacelike hypersurface has the same two dimensional boundary $S$.

Given a three form $D$, Stoke's theorem tells us that
\begin{equation}
 \int_\mathscr{V} dD = \int_{\Sigma'} D - \int_{\Sigma} D .
\end{equation}
In our case we take, for each Killing vector $K_{\underline{c}}^b$,
the three form $D_{\underline{c}\,abc} = T^d_{\;\;e} \; K_{\underline{c}}^e \epsilon_{abcd}$;
so that 
$dD_{\underline{c}\,abcd} = k \nabla_f \left(T^f_{\;\;e} \; K_{\underline{c}}^e \right)\epsilon_{abcd}$;
where $k$ is a constant.

Therefore our main equation is
\begin{equation}\label{eq:dD}
 0 = \int_\mathscr{V} dD_{\underline{c}} = \int_{\Sigma'} D_{\underline{c}} - \int_{\Sigma} D_{\underline{c}} ;
\end{equation}
since due to the conservation equation (\ref{eq:conservT2}) the left hand side vanishes.
It is because of this reason that actually the total momentum is 
determined by $S$, and not by the particular hypersurface with boundary $S$.

This equation is intimately related to the equations of motion;
since equations (\ref{eq:dD}) are telling us that the difference
of the total momentum calculated at $\Sigma'$ and $\Sigma$ vanishes.
We will apply these equations to the case of a particle, even if one
has indications that each term in the difference could be ill defined;
since it could contain infinite terms; however its difference is finite.

Let $S$ be a sphere at future null infinity defined as the asymptotic sphere
of the future null cone of a point $Q(\tau)$; and let $\Sigma$
be the future null cone of this point.
\begin{figure}[htbp]
\centering
\includegraphics[clip,width=0.5\textwidth]{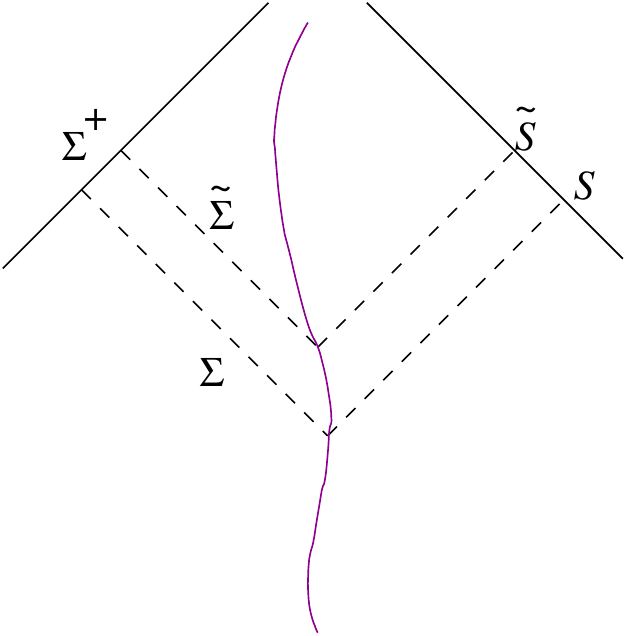}
\caption{Two hypersurfaces reaching future null infinity.
}
\label{fig:scri1}
\end{figure}
Let $\tilde S$ be the corresponding asymptotic sphere for the point $Q(\tau+d\tau)$
on the curve $C$; and let $\tilde\Sigma$
be the future null cone of this point.
Let us call $\Sigma^+$ the hypersurface at future null infinity with boundaries
$S$ and $\tilde S$.
Then in the last equation we can identify $\Sigma'= \tilde\Sigma \cup \Sigma^+$.
See Fig. \ref{fig:scri1}  for a graphical representation of the hypersurfaces.

Therefore, assuming the same integrand, one has the relation
\begin{equation}\label{eq:simgas}
\int_{\tilde\Sigma}
-
\int_\Sigma
=
-\int_{\Sigma^+}
;
\end{equation}
which says that the momentum at $\tilde S$ is the momentum at $S$,
minus the flux through ${\Sigma^+}$.

Let us differentiate between a charged particle, which we will call
the system of interest $A$, and the rest of the world which we will
call system $B$. Then, the total energy momentum tensor can be decompose
as the following sum of terms
\begin{equation}\label{eq:Tdecomposition}
 T = T_{\text{(m)}A} + T_{\text{(m)}B} +
T_{\text{(EM)}A} + T_{\text{(EM)}B} + T_{\text{(EM)}A.B} ;
\end{equation}
where we have distinguished: the mechanical term (m) of particle $A$,
the mechanical term of system $B$,
the electromagnetic term (EM) of particle $A$,
the electromagnetic term of system $B$,
the electromagnetic term of the products of fields of particle $A$
times fields of system $B$.

Considering the vector $T^a_{\,b} K_{\underline{c}}^b$;
whose divergence is zero, one can apply equation (\ref{eq:simgas}).
Each term of (\ref{eq:Tdecomposition}) contributes on the left hand side
with a corresponding $dP$ term, 
i.e. a difference of momentum at both null hypersurfaces.
The right hand side can be expressed as
\begin{equation}
 -\int_{\Sigma^+} = -\int_{S} d\tau ;
\end{equation}
where $S$ is the intersection of the null cone $\Sigma$ with future null 
infinity, 
and we are considering an infinitesimal $d\tau$.

Therefore one has
\begin{equation}\label{eq:dPdecomposition}
\begin{split}
 d&P_{\text{(m)}A} + dP_{\text{(m)}B} +
dP_{\text{(EM)}A} + dP_{\text{(EM)}B} + \\
& dP_{\text{(EM)}A.B} + dP_{\text{(EM)}B.A}
=
-\int_{S} T_{ab} K^b n^a \frac{dS^2}{V^2} d\tau
;
\end{split}
\end{equation}
where $K^b$ is a translational Killing vector, $n^a$ is the null vector normal to future null infinity
satisfying $n^a l_a= 1$, and $dS^2$ is the surface element of the unit sphere.
We have also indicated separately the difference in momentum of system $A$ due
to the existence of system $B$ $(dP_{\text{(EM)}A.B})$, and 
the difference in momentum of system $B$ due
to the existence of system $A$ $(dP_{\text{(EM)}B.A})$.

For the sake of simplicity in the discussion, let us consider first the case
en which system $B$ consists of a smooth distribution of matter and charges
with no radiation at future null infinity.
Furthermore, let us assume that system $B$ can be represented by a Lagrangian
formulation, included the actions of system $A$ on $B$.
Therefore, under variations of the $B$ fields, the Lagrangian will
induce the equations of motion
\begin{equation}
\frac{dP_{\text{(m)}B}}{d\tau} 
+ \frac{dP_{\text{(EM)}B}}{d\tau} 
+ \frac{dP_{\text{(EM)}B.A}}{d\tau} = 0
;
\end{equation}
in other words, the terms 
$dP_{\text{(m)}B} + dP_{\text{(EM)}B} + dP_{\text{(EM)}B.A}$
are balanced in equation (\ref{eq:dPdecomposition}),
which implies
\begin{equation}\label{eq:dPbalance1}
\frac{dP_{\text{(m)}A}}{d\tau} 
+ \frac{dP_{\text{(EM)}A}}{d\tau} 
=
- \frac{dP_{\text{(EM)}A,B}}{d\tau} 
-\int_{S} T_{ab} K^b n^a \frac{dS^2}{V^2} 
.
\end{equation}

Then one can think that the term $dP_{\text{(EM)}A}$ could contain
infinite contributions at the particle world line $C$, 
since the self fields will be taken into account.
However we know that the right hand side of the equation is finite;
therefore, the sum of both terms on the left is finite.

To calculate the flux term it is convenient to note that
\begin{equation}
 \frac{1}{4 \pi} \int l^a \frac{dS^2}{V^2}
 = v^a ,
\end{equation}
\begin{equation}
 \frac{1}{4 \pi} \int l^a l^b \frac{dS^2}{V^2} = \frac{4}{3} v^a v^b - \frac{1}{3} \eta^{ab} 
\end{equation}
and
\begin{equation}\label{eq:3ls}
 \frac{1}{4 \pi} \int l^a l^b l^c \frac{dS^2}{V^2} = 2 v^a v^b v^c 
- \frac{1}{3} \left( \eta^{ab} v^c + \eta^{bc} v^a + \eta^{ca} v^b \right) 
;
\end{equation}
which are a generalization of the relations that appear in \cite{Poisson:1999tv}.

The radiative part of $T_{ab}$ is given by
\begin{equation}
 t_{ab} = \frac{1}{2\pi} \phi_2 \bar\phi_2 l_a l_b ;
\end{equation}
so that the integration on ${S}$ is more precisely
\begin{equation}
\begin{split}\label{eq:flux2}
 \int_{S} \frac{1}{2\pi} \phi_2 \bar\phi_2 l_{\underline{c}} \frac{dS^2}{V^2} 
&=
 \frac{e^2}{2\pi}\int_{S}  V^2 
\eth\left( \frac{\dot V}{V}\right) \bar\eth\left( \frac{\dot V}{V}\right)
l_{\underline{c}} 
\frac{dS^2}{V^2}  \\
&=
 \frac{e^2}{2\pi}\int_{S}  
\eth_V\left( \frac{\dot V}{V}\right) \bar\eth_V\left( \frac{\dot V}{V}\right)
l_{\underline{c}} 
\frac{dS^2}{V^2} ;
\end{split}
\end{equation}
where $\eth_V$ is the edth\cite{Geroch73} operator for the metric with surface element $\frac{dS^2}{V^2}$.

Now, let us note that for any two vectors $A^a$ and $B^b$, and defining
the quantities $A= A^a l_a$ and $B= B^a l_a$, one has
\begin{equation}
 A^a B_ a = A B 
+ A \bar\eth_V \eth_V B + B \bar\eth_V \eth_V A 
- \eth_V A \bar \eth_V B - \bar\eth_V A \eth_V B ;
\end{equation}
where we are assuming the Minkowski metric in the contraction of vectors.

Therefore, noting that $\frac{\dot V}{V} = \dot v^a l_a$, one deduces that
\begin{equation}
\begin{split}
 \eth_V\left( \frac{\dot V}{V}\right) \bar\eth_V\left( \frac{\dot V}{V}\right)
&=
\frac{1}{2}\left( \frac{\dot V}{V}\right)^2 + \frac{\dot V}{V} \bar\eth_V \eth_V\frac{\dot V}{V}
- \frac{1}{2}\dot v^a \dot v_a \\
&=
 - \frac{1}{2}\left( \frac{\dot V}{V}\right)^2 
- \frac{1}{2}\dot v^a \dot v_a \\
&=
- \frac{1}{2} \left( \dot v^a l_a \dot v^b l_b + \dot v^a \dot v_a \right) 
;
\end{split}
\end{equation}
where we have used that
\begin{equation}
 \bar\eth_V \eth_V\frac{\dot V}{V} = -\frac{\dot V}{V} .
\end{equation}

Then, equation (\ref{eq:flux2}) becomes
\begin{equation}
\begin{split}\label{eq:flux3}
 \int_{S} \frac{1}{2\pi} \phi_2 \bar\phi_2 l_{\underline{c}} \frac{dS^2}{V^2} 
&=
 \frac{e^2}{2\pi}\int_{S}  
\eth_V\left( \frac{\dot V}{V}\right) \bar\eth_V\left( \frac{\dot V}{V}\right)
l_{\underline{c}} 
\frac{dS^2}{V^2} \\
&= 
- \frac{e^2}{4\pi}\int_{S}  
\left( \dot v^a l_a \dot v^b l_b
+ \dot v^a \dot v_a
\right) l_{\underline{c}} 
\frac{dS^2}{V^2} \\
&= 
- e^2
\left[ \left( -\frac{1}{3}
\dot v^a \dot v^b \eta_{ab}v_{\underline{c}}
+  \dot v^a \dot v_a  v_{\underline{c}}
\right)
\right] \\
&= 
- \frac{2}{3} e^2 
 \dot v^a \dot v_a  v_{\underline{c}}
.
\end{split}
\end{equation}

Finally, the flux term contributes with
\begin{equation}
 -\int_{S} T_{ab} K_{\underline{c}}^b n^a \frac{dS^2}{V^2} 
=
\frac{2}{3} e^2 
 \dot v^a \dot v_a  v_{\underline{c}} .
\end{equation}

The other term on the right hand side of (\ref{eq:dPbalance1})
includes all the contributions due to forces that system $B$ exerts
on particle $A$; for the case of electromagnetic interactions
one has
\begin{equation}
 - \frac{dP_{\text{(EM)}A,B\,{\underline{c}} } }{d\tau} 
=
-e F(B)_{a{\underline{c}}} v^a 
=
e F(B)_{{\underline{c}}a } v^a 
;
\end{equation}
which is the standard Lorentz force.
In order to elucidate the role of units, we clarify that we are using units of time for $\tau$
(the proper time);
so that a dot, or proper time derivative, implies an extra 1/second in any expression.

The other terms on the left hand side of (\ref{eq:dPbalance1}),
not only include the mechanical term proportional to the acceleration 
$\dot v_{\underline{c}}$, but they also include other terms
coming from the contributions due to the self fields of the particle.
The main idea in our presentation is not to treat both terms separately but
as a unity; namely, the variation of the total momentum.
In the customary treatment in which the terms are treated separately, one 
has to deal with infinities and complicated arguments for their cancellation.
We show here that the variation of the total momentum does not include difficulties.

The variation of the total momentum does
 not have arbitrary vectorial dependence;
since to begin with, the mechanical contribution only depends on the four velocity, and
the energy momentum tensor depends on the electromagnetic tensor,
and this depends on the velocity and acceleration.
Therefore, the time derivative of the total momentum 
can only depend at most on $(v^c, \dot v^c,\ddot v^c)$; in other words it can not depend
on higher derivatives like $(\dddot v^c)$.

The coefficients of this vectors must be finite, since the right
hand side is finite.
Therefore, although at first sight one could expect divergent behavior
for the expressions of the self fields; one must understand that
there are cancellation of infinite like terms between the mechanical
and the field momenta, that as a result provide with the finite
coefficients.

In other words, we do not attempt to decompose the variation of total momentum
into a mechanical and electromagnetic part, but consider an expression for
the variation of the total momentum; which we know is finite, and of the form
\begin{equation}\label{eq:pdot1}
\dot P_{\underline{c}}=
m' v_{\underline{c}} + m \, \dot v_{\underline{c}} 
+ \alpha \, \ddot v_{\underline{c}},
\end{equation}
were $m',m$ and $\alpha$ are understood as functions of $\tau$.

From equation (\ref{eq:dPbalance1}) we conclude then that the equations of motion can be written as
\begin{equation}\label{eq:balance2}
m' v_{\underline{c}} + m  \dot v_{\underline{c}} 
+ \alpha \ddot v_{\underline{c}}
=
e F(B)_{{\underline{c}}a } v^a 
+
\frac{2}{3} e^2 
 \dot v^a \dot v_a  v_{\underline{c}} .
\end{equation}

Let us also note that the total momentum must be expressed as
\begin{equation}\label{eq:ptotal}
P_{\underline{c}}=Mv_{\underline{c}} + \alpha  \dot v_{\underline{c}}.
\end{equation}
with $M=M(\tau)$.
This expression for the total momentum is a generalization of the expression (4.4)
found in reference \cite{Teitelboim70} through renormalization procedures.

By taking the time derivative of the last equation and comparing with eq.(\ref{eq:pdot1}), we get
\begin{eqnarray}
\dot M&=&m',\\
M+\dot \alpha&=&m . \label{eq:Mmasalfa}
\end{eqnarray}

Let us note that there are at least two notions of mass. One has the parameters $m$,
which we will call \emph{inertial mass}. Also one has the parameter $M$,
that we will call \emph{rest mass}, in order to differentiate them.

Therefore the final form of the equations of motion is
\begin{equation}\label{eq:balance3}
\boxed{
\dot M v_{\underline{c}} + (M +\dot \alpha)  \dot v_{\underline{c}} 
+ \alpha \ddot v_{\underline{c}}
=
e F(B)_{{\underline{c}}a } v^a 
+
\frac{2}{3} e^2 
 \dot v^a \dot v_a  v_{\underline{c}} 
}.
\end{equation}

This is our main result, the most general equations of motion for charged particles in the framework we have presented;
in which it is required the balance of radiated momentum.
Let us note that in our approach we have avoided treating explicitly the elimination of 
infinite renormalizations
or similar techniques.

We recognize new degrees of freedom present in this form of the equations of motion, namely the scalar
$\alpha(\tau)$; and also $M(\tau)$. The latter has been consider already by other authors(see below).

Now, if we contract the expression eq.(\ref{eq:balance3}) with the vector $v^{\underline{c}}$,
and using the notation $-a^2 \equiv  \dot v^a \dot v_a$, we obtain
\begin{equation}\label{eq:mdotalpha}
\dot M
+ \alpha a^2
=
- \frac{2}{3} e^2  a^2
.
\end{equation}
Since this is one equation for two unknown, one must prescribe $\alpha(\tau)$  or $M(\tau)$,
or a relation among them, as we will do below. 

It is interesting to note that if we replace $\dot M$ in the last equation into (\ref{eq:balance3}) 
one obtains
\begin{equation}\label{eq:balance4}
\boxed{
(M +\dot \alpha)  \dot v_{\underline{c}} 
=
e F(B)_{{\underline{c}}a } v^a 
- \alpha (\ddot v_{\underline{c}} - a^2 v_{\underline{c}}  )
};
\end{equation}
which only sets the dynamics for three degrees of freedom since it is orthogonal to the four velocity $v$.
Therefore one has the option to consider the original equations (\ref{eq:balance3}), or 
the two independent equations (\ref{eq:balance4}) and (\ref{eq:mdotalpha}).

In presenting equations (\ref{eq:balance3}), (\ref{eq:mdotalpha}) and (\ref{eq:balance4}) we have
given preference to the rest mass $M$; alternatively, if one gives preference to the inertial mass $m$
one can reexpress them as
\begin{equation}\label{eq:balance3m}
\boxed{
(\dot m - \ddot \alpha) v_{\underline{c}} + m  \dot v_{\underline{c}} 
+ \alpha \ddot v_{\underline{c}}
=
e F(B)_{{\underline{c}}a } v^a 
+
\frac{2}{3} e^2 
 \dot v^a \dot v_a  v_{\underline{c}} 
}.
\end{equation}
\begin{equation}\label{eq:mddotalpha}
\dot m - \ddot \alpha
+ \alpha a^2
=
- \frac{2}{3} e^2  a^2
,
\end{equation}
and
\begin{equation}\label{eq:balance4m}
\boxed{
m  \dot v_{\underline{c}} 
=
e F(B)_{{\underline{c}}a } v^a 
- \alpha (\ddot v_{\underline{c}} - a^2 v_{\underline{c}}  )
}.
\end{equation}

In summary, we have arrived at the general equations of motion for a charged particle, in which
two new degrees of freedom appear, and also the order of the equations for $v$ has changed; from
the Lorentz equations. 
So, the set of basic variables, can be considered
$(v^a,M,\alpha)$ or $(v^a,m,\alpha)$; where the velocity vector is in turn expressed in terms
of the position of the particle.
Since $v$ is always assumed to have unit modulus, one has four equations for five degrees of freedom.
Therefore one has the liberty to choose a relation among the two new degrees of freedom.
We will discuss below how different choices conduces us to known cases and 
also to new physically interesting ones.

It is probably worthwhile to emphasize that we have arrived at these general equations of motion
by balancing the retarded radiation field of the charged particle when the rest of the system
does not radiate. This is a very strong assumption, on the rest of the electromagnetic system,
and in particular would exclude the interaction of the particle with perfect conductors, 
or a dispersive permeable medium, for example.
Our approach is a technique valuable to calculate the `corrections' to the Lorentz force that
take into account the retarded radiation fields; however, after one has calculated such corrections
one expects these to be valid in a general situation.

\subsection{Historical choices on the general equations of motion} 

\subsubsection{The Lorentz-Dirac equations}
We have seen before that there are two natural notions of mass that appear in our treatment.
Therefore it is interesting to consider the particular cases in which each of them is required to
be a constant of the motion; since in particular real elementary particles as the electron
seem to have a constant mass.

The case $\dot m=0$ will be considered below; we will consider here the case $\dot M=0$.

This particular choice can be understood as follows.
From (\ref{eq:ptotal}), we see that $M=m-\dot\alpha$ can be interpreted as the 
rest mass of the electron (which has contributions of the electromagnetic field generated 
by the particle).  Then if we want to describe particles whose total rest mass remains 
constant, we must require that $\dot M=\dot m-\ddot\alpha=0$, then from 
eq.(\ref{eq:mdotalpha}), we obtain that
\begin{eqnarray}
 \alpha &=& - \frac{2}{3} e^2  .
\end{eqnarray}
Furthermore, one can also
deduce from equation (\ref{eq:Mmasalfa}) that the two notions of mass coincide, and therefore
one also has $\dot{m}=0$.

Therefore, one arrives at the following equations of motion for the charge
given by
\begin{equation}\label{eq:lor-dirac}
 m \dot v^a = e F(B)^a_{\;b} v^b 
+\frac{2}{3} e^2  \left( \ddot v^a + \dot v^b \dot v_b  v^a \right) ;
\end{equation}
which are the well known Lorentz-Dirac equations of motion for the electron.

It is probably worth while to recall that these equations have several problems,
as it has been investigated in the past by several authors.
First of all they are  third order differential equations for the position of the
charged particle; which is contrary to the general accepted idea of mechanics
for particle. Secondly, they have the so called problem of the runaway solutions.
Several of the implied problems were discussed by Dirac and we will not review them here.

\subsubsection{The Bonnor equations}
Another approach to the two mass question is to require that the two notions coincide.
Then, let us consider the choice $M=m$. From this one immediately obtains that $\dot \alpha=0$;
and we recognize that the choice $\alpha = - \frac{2}{3} e^2$ coincides with the previous case;
but instead we study here the general $\dot m \neq 0$ case; which must satisfy
\begin{equation}
 \dot m = - \left (\frac{2}{3} e^2 + \alpha\right) a^2 ;
\end{equation}
where as noted before, $\alpha$ must be a constant.

 In particular if we set $\alpha=0$ 
one arrives at the  equations of motion with varying mass, namely
\begin{equation}\label{eq:bonnor}
m  \dot v_{\underline{c}} 
=
e F(B)_{{\underline{c}}a } v^a, 
\end{equation}
with
 \begin{equation}\label{eq:bonnormass}
\dot m 
=
-
\frac{2}{3} e^2  a^2
.
\end{equation}
This equation was studied by \cite{Bonnor74}.
Unfortunately it has the unphysical consequence
that the mass of the particle could vanish in a finite time.

\section{Study of the general equations of motion for charged particles}\label{sec:generalsolutions}

\subsection{Behavior of the general solution for two particular cases}
 
We study here general properties of the solutions that can be deduced prior to the use of the
extra liberty for the choice of a condition among the two new degrees of freedom.

\paragraph{The case $a^2=0$.}

Let us consider first the case in which $a^2=0$ for all times. In this case, 
 all the terms in equations (\ref{eq:balance3}) collapses to zero, and the particle moves along
a geodesic. 
One can observe that the terms involving $\dot v$ and $\ddot v$ are zero; then the only two remaining
terms containing the external fields and the first one proportional to $v$ are orthogonal, so that they
must vanish independently.
In particular one has $F(B)=0$ and  $\dot M=0$.

\paragraph{The case $F(B)=0$.}

The other case that it is important to be considered is the case in which the external fields $F(B)$ are zero,
and study the general behavior of $a^2$ for large values of the time variable; in order to see the 
nature of the runaway solutions problem in this setting.
So now we invert the logic and think what is the behavior of $a^2$ in terms of the behavior
of the new degrees of freedom $M$ and $\alpha$.
We observe then that by contracting the equations of motion (\ref{eq:balance3}) with $\dot v$ one obtains
\begin{equation}\label{eq:a2dot}
  (M + \dot \alpha) a^2 + \frac{\alpha}{2} \dot{ a^2} = 0.
\end{equation}
This constitutes a simple first order differential equation for $a^2$; whose behavior is determined
by the sign of the coefficients.
In this setting a runaway behavior implies that the coefficients have opposite sign.
Since one can choose one relation for the two extra degrees of freedom,
we would like to explore possible conditions for them.
If one were interested in an asymptotic nonincreasing solution for $a^2$, one would probably consider
$(M + \dot \alpha) > 0 $ and $\alpha > 0$.
It is interesting to note that if one chooses as initial data for $\alpha$ and $\dot \alpha$ 
a tiny positive value for them,
then equation (\ref{eq:a2dot}) would imply an initial fast exponential decay of $a^2$.
This, will in turn imply, that initially $M \sim m$, and from (\ref{eq:mdotalpha}) 
that $\dot M$ would become very small in a short time.
This choice of very small  initial conditions of the $\alpha$ degree of freedom
avoid the problem of the runaway solutions, for the case in which $F(B)=0$ for all times.
Instead, the discussion becomes more complicated if one thinks in the situation in which the
external field is turned on and off. When the external field is turned on, one can arrange so that
$\dot{ a^2} > 0$; which will imply from (\ref{eq:a2dot}) that $\alpha$ rapidly adopt a negative
value(provided one has manages to maintain $m=M+\dot \alpha > 0$). If later the external field is turned off again, then the particle will enter the region
of vanishing field with the wrong initial condition and one would probably be in the presence of a 
runaway solution. 
So there is no universal choice of the initial conditions of the $\alpha$ degree of freedom
that would exclude the runaway solutions.
Also, the behavior $\alpha > 0$ would imply, from (\ref{eq:mdotalpha}),
 a condition $\dot M < 0$,
which could have the problem that $M$ would vanish in a finite time, if $a^2$ where bounded from below
by a nonzero value. 
From this analysis we can not exclude the possibility of a solution where both $\alpha$ and
$a^2$ go asymptotically to zero, for large values of the time coordinate, but with a nonzero
$M$ value.
The other possibility is to have $\alpha < 0$; 
but would conduce us to the runaway solution problem(provided one has manages to maintain $m=M+\dot \alpha > 0$). 
Therefore, we conclude that  the 
runaway solution problem is generic in this dynamics too.
The case $(M + \dot \alpha) < 0 $ is ruled out since this would imply that the inertial mass is negative.
The value $\alpha=0$ was not considered due to the fact that it would collapse to the previously
studied case of $a^2=0$ (since we are examining the case $F(B)=0$).

\subsection{Behavior of the solution with $\dot m=0$}

Let us recall that $m$ can be understood as the inertial mass.
The reason that condition $\dot m=0$ is worth studying comes also from the following considerations. 
The time derivative of the \emph{rest mass}\footnote{Recall that $M$ is the factor of $v$ in the 
expression for the total momentum.} 
of the particle is given 
by $\dot M=\dot m-\ddot \alpha$. Therefore, the condition $\dot m=0$ means that 
the inertial mass is constant, and the variation 
in the rest mass is only due to the electromagnetic fields contribution to the mass. 
Note that the Lorentz-Dirac equations are contained in this family; as it is case $F(B)=0$
considered in the last subsection.

Treating the equations of motion as exact to any order leads one to possible divergent behavior
for the new degrees of freedom as we will see next.

In this case,
equation (\ref{eq:mdotalpha}) can be completely expressed in terms of $\alpha$; which must satisfy 
\begin{equation}\label{eq:mdotcero}
\ddot \alpha -a^2\, \alpha 
=
\frac{2}{3} e^2  a^2   .
\end{equation}
It is convenient to define $\beta=\alpha + \frac{2}{3} e^2$, since then
 eq.(\ref{eq:mdotcero}), can be written as
\begin{equation}\label{eq:beta}
\ddot \beta-a^2\, \beta=0 .
\end{equation}

In appendix \ref{ap:beta} there is a brief discussion of the properties of solutions
of equation (\ref{eq:beta}) in terms of the global properties of $a^2(\tau)$.
There, a list of theorems is presented where nondivergent properties for $a^2$
are assumed for large values of $\tau$.
These results suggest that one could in principle be able to choose initial conditions
for $\beta$ to select decaying solutions. That is, one in general would find
independent solutions for $\beta$; one that would grow with time and other that would
decay to zero. It is somehow remarkable that one could select a solution for $\beta$
so that $\alpha$ tends asymptotically to the Lorentz-Dirac value.
However, we emphasize that in order to find the decreasing solution one must know
the whole future history of the motion of the particle; therefore this is a nonlocal
analysis of the motion; which in particular is not based just on initial conditions.

Let us note that defining the vector $f^a = e F(B)^a_{\;b } v^b$, equations (\ref{eq:balance4m}) become
\begin{equation}\label{eq:cuasiLorentzDirac}
 m \dot v^a = f^a - \alpha (\ddot v^a - a^2 v^a) ;
\end{equation}
which with (\ref{eq:mdotcero})
constitute two independent equations.
Equation (\ref{eq:cuasiLorentzDirac}) has the same form as the Lorentz-Dirac equations;
and it would agree with it if one would take the solution $\alpha =- \frac{2}{3}e^2$, as
mentioned above.

There is a general concern with the dynamical equations chich involves the choice of two constants
for the initial conditions for equation (\ref{eq:mdotcero}). Then in turn one would have the 
set of solutions of the equations of motion (\ref{eq:balance3}) affected by the arbitrary 
initial choice for the $\alpha$ degree of freedom.

\section{Dynamics in terms of orders of the strength of the interaction with external fields}\label{sec:orders}

The difficulties found in the study of the general solutions, of the last section,
suggest that the original equations of motion must be understood in terms
of an analysis in orders of the interaction with the external fields, which we present next.

\subsection{Equation of motion up to second order in the external fields}

It is important to remark that equations (\ref{eq:balance3}) are  exact equations
in the classical framework for charged particles.
It is generally believed that a classical description of particles must break down when one
considers real elementary particles at microscopic levels.
Presumably, one would be able to describe the behavior of the new degrees of freedom from
quantum electrodynamics.
All this suggests that equations (\ref{eq:balance3}) should be understood in terms of orders
of the interaction with the external field $F(B)$.

Therefore next, we present a study of our main equations (\ref{eq:balance3}) in first orders 
of the strength of the interaction with the external field; but still considering the physical condition $\dot m = 0$.
This condition allows to interpret $M(\tau)$ in terms of $\alpha(\tau)$.

Note that contracting $f^a$ with eq. (\ref{eq:balance3}),
one obtains
\begin{equation}\label{eq:fbalance}
 m f^a   v_a
+ \alpha f^a  \ddot v_a
=
- {\bf f}^2
,
\end{equation}
with ${\bf f}^{2} \equiv - f_af^a$.
At this point it is convenient to recall that the equations of motion for a test particle
with negligible mass and charge $\mu$ and $q$, respectively, is
\begin{equation}\label{eq:lorentz}
\mu  \dot v_{\underline{c}} 
=
q F(B)_{{\underline{c}}a } v^a 
;
\end{equation}
from which one deduces that the behavior of $\alpha(\tau)$ is completely
due to the interaction of the charge $e$ with the external field $F_{ab}$.
If one considers the correction to the equations of motion as arising from the strength  of different terms
appearing in eq.(\ref{eq:balance3}), one is tempted to consider terms of order $\mathscr{O}({\bf f}^p)$
and also to consider order in $\mathscr{O}(e^q)$; since the $e^2$ appears as an independent factor
in the radiation term.

Let us note that from equations (\ref{eq:lorentz}) one can deduce that $\dot v = \mathscr{O}({\bf f})$.
Then, since we know that $\alpha = \mathscr{O}({\bf f}^0)$; one would have, from equation (\ref{eq:fbalance}) 
that $\ddot v = \mathscr{O}(\dot {\bf f})=\mathscr{O}( {\bf f})$. 
Then, since $\alpha$ is expected to have some nontrivial $\mathscr{O}(e^q)$ order, the second term
on the left hand side of (\ref{eq:fbalance}) is of higher order than the rest;
as can also be deduced from the  fact that 
\begin{equation}
 \dot v = \frac{1}{m} f + \mathscr{O}({\bf f}^{+}) .
\end{equation}
where $\mathscr{O}({\bf f}^{+})$ means higher order than $({\bf f})$,
as for example $\mathscr{O}(e^2{\bf f})$, and for simplicity we have omitted the vectorial index.

Now we will consider  equations (\ref{eq:cuasiLorentzDirac}) up to order $\mathscr{O}({\bf f}^2)$. 
We therefore study equation (\ref{eq:beta}) up to order $\mathscr{O}({\bf f}^2)$.
For this purpose we will use $a=\frac{{\bf f}}{m}$ and suggest a $\beta$ of the form
\begin{equation}
 \label{eq:beta2}
\beta(\tau) = A_0(\tau) + A_1(\tau) {\bf f} + A_2(\tau) {\bf f}^2 +  \mathscr{O}({\bf f}^3) .
\end{equation}

Then equating order like terms in equation (\ref{eq:beta}) one has:
\begin{equation}\label{eq:aya0}
   \ddot A_0   
= 0 ,
\end{equation}
\begin{equation}\label{eq:aya1}
  \ddot {(A_1 {\bf f})}
= 0 ,
\end{equation}
and
\begin{equation}\label{eq:aya2}
  \ddot {(A_2 {\bf f}^2)} 
 = A_0 \frac{{\bf f}^2}{m^2}
 .
\end{equation}
The solution of equation (\ref{eq:aya1}) implies two constants of integration that must
be of $\mathscr{O}({\bf f})$; but for the general physical situation of a nonstationary ${\bf f}$
we find no universal way to assign a constant. Therefore we chose those two constants to be zero;
in other words, we set $A_1 {\bf f}= 0$.

Instead in the solution of (\ref{eq:aya0}) one finds $A_0 = A_{00} + A_{01} \tau $; where
$A_{00}$ and $A_{01}$ are constants.

Replacing the solution of these terns in equations (\ref{eq:cuasiLorentzDirac}) and keeping terms up to 
$\mathscr{O}({\bf f}^2)$ one obtains
\begin{equation}\label{eq:balance3-2b}
 m  \dot v^a
=
f^a
+ (  \frac{2}{3} e^2  - A_0 )
\left(
\frac{1}{m }\dot f^a -  \frac{{\bf f}^2}{m^2} v^a
\right)
.
\end{equation}
It is hard to give a physical meaning to a dynamical system that depend on the two
parameters involved in $A_0$. 
Even if one chooses $A_{01}=0$ to avoid the time dependence; one is still left with a constant
that probably should be determined from another theoretical framework like quantum electrodynamics. 
From the classical point of view one could consider studying the cases in which $A_0$ is taken
to be a constant proportional to $e^2$. A peculiar case would be to consider $A_0= \frac{2}{3}e^2$;
since this choice would just cancel the other terms proportional to $e^2$, and therefore
conduce to the Lorentz force for the charged particle. This would be unacceptable in our
approach to the equations of motion that takes into account the change in momentum due to 
the radiation emitted by the particle.
However, considering a background quantum nature of particles, one could think in a value of the form
\begin{equation}
 A_0 = \lambda \frac{h}{c^2};
\end{equation}
where $h$ is Planck constant, $c$ the velocity of light and $\lambda$ a number without units that
should be determined from quantum electrodynamics.
Then, since we are considering second order expressions, in the equations of motion (\ref{eq:balance3-2b}),
in place of $\dot f_{\underline{c}}$ one must use
\begin{equation}
\begin{split}
 \dot f_{(2)\underline{c}} =& \dot{(e F(B)_{\underline{c}\,b }\; v^b)}
= e\dot{(F(B)_{\underline{c}\,b })}\; v^b + e F(B)_{\underline{c}\,b }\; \dot{v}^b \\
=& e\dot{(F(B)_{\underline{c}\,b })}\; v^b + \frac{e}{m} F(B)_{\underline{c}\,b }\; f^b 
.
\end{split}
\end{equation}

In a complete classical framework, one would not have any physical argument for a nonzero $A_0$;
and therefore one would be forced to take $A_0=0$; leading to the equations of motion
\begin{equation}\label{eq:balance3-20}
 m  \dot v^a
=
f^a
+ \frac{2}{3} e^2  
\frac{1}{m }\dot f_{(2)}^a
-
\frac{2}{3} e^2
 \frac{{\bf f}^2}{m^2} v^a
;
\end{equation}
which coincides with the Landau-Lifshitz\cite{Landau75} equations of motion.

\subsection{Study of the equations of motion up to third order in the interaction with external fields}

Now we would like to study equations (\ref{eq:balance3}), or equivalently (\ref{eq:balance4m}) up to third
order.

Let us note that by having obtained the second order equations of motion in the previous section;
one has a second order acceleration $\dot v_{(2)}$; 
namely, the one that satisfies equations (\ref{eq:balance3-20}). 
Then, third order in turn is defined to satisfy
the equations of motion
\begin{equation}\label{eq:balance4m-3ord}
\boxed{
m  \dot v_{(3)}^b
=
f^b
- \alpha_{(2)} (\ddot v_{(2)}^b - a_{(2)}^2 v^b  )
}
;
\end{equation}
where $\ddot v_{(2)}^b$ is the $\tau$ derivative of
\begin{equation}
 \dot v_{(2)}^b =  
\frac{1}{m } f^b
+   (  \frac{2}{3} e^2  -  A_0 )
\frac{1}{m^2 }\dot f_{(2)}^b
-
(  \frac{2}{3} e^2  -  A_0 )
 \frac{{\bf f}^2}{m^3}
v^b 
;
\end{equation}
and one should consider terms up to order 3.

In this way one can see that this procedure can be generalized to any higher order where
$\dot v_{(n)}^b$ satisfies the differential equations
\begin{equation}\label{eq:balance4m-nord}
\boxed{
m  \dot v_{(n)}^b 
=
f^b
- \alpha_{(n-1)} (\ddot v_{(n-1)}^b - a_{(n-1)}^2 v^b  )
}
;
\end{equation}
where $\ddot v_{(n-1)}^b$ is the time derivative of
\begin{equation}\label{eq:balance4m-n-1-ord}
  \dot v_{(n-1)} ^b
=\frac{1}{m }
f^b
- \alpha_{(n-2)} (\ddot v_{(n-2)}^b - a_{(n-2)}^2 v^b  )
;
\end{equation}
and one should only consider terms up to order $n$.

Then, coming back to the third order calculation, let us consider $\beta$ of the form
\begin{equation}
 \label{eq:beta3}
\beta(\tau) = A_0(\tau) + A_1(\tau) {\bf f} + A_2(\tau) {\bf f}^2 +  A_3(\tau) {\bf f}^3 + \mathscr{O}({\bf f}^4) .
\end{equation}
Then equating order like terms in equation (\ref{eq:beta}) one obtains equations
(\ref{eq:aya0}) and (\ref{eq:aya1}); while instead of (\ref{eq:aya2})  one now has
\begin{equation}\label{eq:aya2b}
  \ddot {(A_2 {\bf f}^2)} 
 = A_0 a_{(2)}^2 
 ,
\end{equation}
and also
\begin{equation}\label{eq:aya3}
  \ddot {(A_3 {\bf f}^3)} 
 = A_0 a_{(3)}^2 
 ;
\end{equation}
where $a_{(2)}^2$ and $a_{(3)}^2$ are the order $\mathscr{O}({\bf f}^2)$ and $\mathscr{O}({\bf f}^3)$
respectively of $a^2$.

If $A_0 \neq 0$, 
the main difficulty in this case is the fact that in the product $\beta a^2$ of equation (\ref{eq:beta})
one needs the explicit integral of (\ref{eq:aya2b}), which will involve a couple of integration constants
(or order $\mathscr{O}({\bf f}^2)$)
associated to the choice of initial time for the integration. 
Since the integration constants must be of order $\mathscr{O}({\bf f}^2)$, they can not be associated
just with the charge `$e$' or to universal constants; therefore it would be very difficult to give 
physical meaning to a physical dynamical system that depends on the arbitrary choice of initial conditions
(even if one manage to choose them of order $\mathscr{O}({\bf f}^2)$),

All this seems to indicate that the only physically sensible choice is to take $A_0 = 0$;
since in this case one would have to solve the homogeneous problem, with the natural choice 
for all arbitrary constants to be zero. 
In other words, $\beta$ must be zero and we are 
conduced to the equations
\begin{equation}\label{eq:balance4m-3ord-b}
\boxed{
m  \dot v_{(3)}^b
=
f^b
+ \frac{2}{3} e^2   (\ddot v_{(2)}^b - a_{(2)}^2 v^b  )
}
;
\end{equation}
where
\begin{equation}\label{eq:dotv2-b}
 \dot v_{(2)}^b =  
\frac{1}{m } f^b
+   \frac{2}{3} e^2 
\frac{1}{m^2 }\dot f_{(2)}^b
-
  \frac{2}{3} e^2  
 \frac{{\bf f}^2}{m^3}
v^b 
;
\end{equation}
in which we have replaced $\alpha_{(2)} = -\frac{2}{3} e^2$
and one should consider terms up to order 3.

This difficulty will appear in higher orders too; from which we conclude that
$\alpha_{(n \geqslant 2)} = -\frac{2}{3} e^2$, and that 
 the second order case
treated previously is sort of peculiar, since it is the only order that allows for a physically
permissible nonzero $\beta$.

\section{Final comments}
The equations of motion (\ref{eq:balance3}) are general equations for charged particles
that are derived from the condition of balance of variation of total momentum with 
the radiated momentum. In this derivation, we have avoided dealing explicitly with infinite
contributions to the momentum.

We have shown how different versions of equations of motion for charged particles can
be obtained from our general equations; in particular the celebrated Lorentz-Dirac equations.

Studying the properties of solutions to the general equations of motion we have found the 
possibility to choose initial conditions for the new degrees of freedom, which select
decaying modes to the Lorentz-Dirac value. However this choice is possible only if one
knows the whole history of the world line of the particle. This is reminiscent of 
the notion of a horizon in general relativity, which it depends on the whole history
of the spacetime.
Even if one manage to chose these preferable initial conditions, one is still left
with the problem of runaway solutions, or equivalently of pre-accelerations.
If one were forced to a theoretical framework based purely on classical considerations,
this would be a sort of dead-end.
Nonetheless, if one uses the known physical information that the ultimate nature
of real particles is of a quantum kind; and therefore classical theoretical frameworks
should be understood as approximate models of the real world, one would think
that one should not demand a real physical interpretation to the exact classical
equations. Instead one is tempted to consider the corrections to the Lorentz force,
as also being related to terms that probably should be calculated from quantum
electrodynamics. If so, then the original equations should be understood in terms
of orders of the strength of the interactions with the external fields.
In particular there is a natural limit to the strength of the interactions that must be
considered in order to avoid the quantum creation of pairs of particles.

In the study in terms of orders of the strength of interaction with the external fields
we have found at second order the equations of motion (\ref{eq:balance3-20}).
These equations have also been supported in the derivation of reference \cite{Gralla:2009md}
based on a different setting; namely, the study of first effects for 
particles with `small' charge and mass. Instead herewe have considered finite charge and
mass particles but study the exact equations of motion in terms of orders of the strength of the
interaction with the external fields. 
Then, it is no surprise that both approaches agree at first orders.

It is important to remark however that the original Landau-Lifshitz equations 
are of second order and coincide with our equations (\ref{eq:balance3-20}).
Our second order equations (\ref{eq:balance3-2b}) could account  for first
order quantum corrections to the classical equations.

Instead,
the equations of motion  (\ref{eq:balance4m-nord}) and (\ref{eq:balance4m-n-1-ord})
presented here, are  a generalization valid up to any desired higher order.

Summarizing; although we have found more general equations of motion for charged particles,
we have shown that our set contains the main cases studied in the past.
We also give arguments that indicate that one should not take the exact general
equations (\ref{eq:balance3}) as the physically relevant ones. 
Instead, to our understanding the only physically reasonable treatment of the equations of  
motion (\ref{eq:balance3}) is through the notion of finite orders in terms of $\mathscr{O}({\bf f})$;
so that we conclude that the equations of motion applicable to 
classical particles, but with finite charge,
are (\ref{eq:balance3-20}), in second order, or equations (\ref{eq:balance4m-nord}) 
and (\ref{eq:balance4m-n-1-ord}) in higher orders.

The final appendix section presents properties of the solutions to equation (\ref{eq:beta}).

\subsection*{Acknowledgments}

We have benefited from discussions with R. Wald at an early stage of preparation of the manuscript,
and we are very grateful to R. Geroch for his numerous suggestions that contributed to the improvement of the
presentation of our work.
We acknowledge support from CONICET and SeCyT-UNC.

\appendix

\section{Properties of $\beta$ function}\label{ap:beta}

Here we would like to recall some of the properties of the solutions to equation (\ref{eq:beta}).

\begin{teo}\footnote{Page 359 in reference \cite{Hille69}.}
 Let $a^2(\tau) > 0$ be continuous in $(-\infty,\infty)$. Then the equation
\begin{equation}\tag{\ref{eq:beta}}
\ddot \beta-a^2\, \beta=0 ,
\end{equation}
has one and only one solution $\beta_1(\tau)$ passing through (0,1) which is positive and
strictly decreasing for all $\tau$ and one and only one solution $\beta_2(\tau)$ through
(0,1) which is positive and strictly increasing for all $\tau$.
Furthermore 
\begin{equation*}
 a \beta_1 \in L_2(0,\infty), \qquad \dot \beta_1 \in L_2(0,\infty).
\end{equation*}
\end{teo}
The notation for $L_2(0,\infty)$ comes from 
the Lebesgue spaces $L_p(b,c)$; which contain the set of integrable functions such that
\begin{equation}
 || f ||_p = \left[ \int_b^c |f|^p d\tau \right]^\frac{1}{p} < \infty .
\end{equation}

\begin{teo}
If $0 < a_{\text{inf}}^2 < a^2 < a_{\text{sup}}^2 < \infty$, 
then the solution $\beta_1$ satisfies
\begin{equation}
 e^{-a_{\text{inf}} \tau } \leqslant  \beta(\tau)  \leqslant e^{-a_{\text{sup}} \tau } ;
\end{equation}
for  $\tau > 0$.
\end{teo}

\begin{teo}\footnote{Page 445 in reference \cite{Hille69}.}
 Let $a^2$ be positive and continuous in $[0,\infty)$ and posses continuous first and second
order derivatives. Set
\begin{equation}
 H = \frac{\frac{5}{16} (\dot {a^2})^2 - \frac{1}{4} a^2 \ddot {a^2} }{a^5} ,
\end{equation}
and suppose that
\begin{equation}
 H \in L_1(0,\infty) .
\end{equation}
Then there exist constants $c_1$ and $c_2$ such that
\begin{equation}
 \beta_1(\tau) = c_1 
\frac{-e^{ \int_0^\tau a(\tau') d\tau' }}{\sqrt{a}}
\left( 1 + R_1(\tau) \right) ,
\end{equation}
\begin{equation}
 \beta_2(\tau) = c_2
\frac{e^{ \int_0^\tau a(\tau') d\tau' }}{\sqrt{a}}
\left( 1 + R_2(\tau) \right) ,
\end{equation}
where
\begin{equation}
 |R_1(\tau)| \leqslant e^{ \int_\tau^\infty |H(\tau')| d\tau'} - 1 ;
\end{equation}
and a positive constant $c$ such that $|R_2(\tau)| \leqslant c |R_1(\tau)|$.

\end{teo}

\begin{teo}\footnote{Page 380 in reference \cite{Hartman64}}
 In the equation 
\begin{equation}\tag{\ref{eq:beta}}
\ddot \beta-a^2\, \beta=0 ,
\end{equation}
let $a^2$ be a continuous complex-valued function for large $\tau$ satisfying
\begin{equation}\label{eq:a2int}
 \int^\infty \tau |a^2(\tau)| d\tau < \infty ,
\end{equation}
or, more generally
\begin{equation*}
\begin{split}
 A(\tau) \equiv& \int_\tau^\infty a^2(\tau') d\tau' 
=
\lim_{T \to \infty} \int_\tau^T a^2(\tau') d\tau' \;\;\;\text{exists and} \\
&\int^\infty \sup_{\tau \leqslant r < \infty} |A(r)| d\tau < \infty .
\end{split}
\end{equation*}
Then, there is a pair of solutions $\beta_0$ and $\beta_1$ satisfying, as $\tau \to \infty$,
\begin{equation}\label{eq:bet0}
 \beta_0(\tau) \sim 1, \qquad \dot \beta_0(\tau) = o\left( \frac{1}{\tau}\right) ,
\end{equation}
\begin{equation}\label{eq:bet1}
 \beta_1(\tau) \sim \tau, \qquad \dot \beta_1(\tau) \sim 1 .
\end{equation}
Conversely, if $a^2(\tau)$ is real-valued and does not change signs and if (\ref{eq:beta}) has
a solution satisfying (\ref{eq:bet0}) or (\ref{eq:bet1}), then (\ref{eq:a2int}) holds.

\end{teo}

\begin{teo}\footnote{Page 381 in reference \cite{Hartman64}}
 In the equation 
\begin{equation}\label{eq:betlambda}
 \ddot \beta - \left( \lambda^2 + q(\tau) \right) \beta = 0 ,
\end{equation}
let $\lambda > 0$ and $q(\tau)$ be a complex-valued continuous function for large $\tau$
satisfying
\begin{equation}\label{eq:modqless}
 \int^\infty |q(\tau)| d\tau < \infty ,
\end{equation}
or, more generally,
\begin{equation}
\begin{split}
 \int^\infty &q(t) e^{-2\lambda t} dt 
= \lim_{T \to \infty} \int^T \text{exists and} \\
&\int^\infty e^{2\lambda\tau} \sup_{\tau \leqslant s <\infty}
\left| \int_s^\infty q(r) e^{-2\lambda r} dr \right| d\tau < \infty .
\end{split}
\end{equation}
Then, (\ref{eq:betlambda}) has solutions $\beta_0$, $\beta_1$ satisfying
\begin{equation}\label{eq:condbeta01}
 \beta_0 \sim - \frac{\dot\beta_0}{\lambda} \sim e^{-\lambda \tau},
\qquad
\beta_1 \sim - \frac{\dot\beta_1}{\lambda} \sim e^{\lambda \tau}.
\end{equation}
Conversely, if $q(\tau)$ is real-valued and does not change signs and if (\ref{eq:betlambda}) has
a solution $\beta_0$ or $\beta_1$ satisfying the corresponding conditions in (\ref{eq:condbeta01}),
then (\ref{eq:modqless}) holds.
\end{teo}

An example of a solution that does not satisfy any of the hypothesis of the theorems is the case 
$a^2 = \frac{2}{\tau^2}$\footnote{We are very grateful to R. Geroch for pointing out this case.}.
The two independent solutions for $\beta$ are proportional to $\tau^2$ and to $1/\tau$.

%

However, let us note that the case $a^2 = \frac{2 }{\tau^2 + \delta^2} + \epsilon^2$,
satisfy the hypothesis of the last theorem; which 
states that there is an exponentially decreasing solution of 
the form $\beta_0 \sim e^{-\epsilon \tau}$.




\begin{thebibliography}{10}
\expandafter\ifx\csname natexlab\endcsname\relax\def\natexlab#1{#1}\fi

\bibitem[Dirac(1938)]{Dirac38}
P.~Dirac, ``Classical theory of radiating electrons'', {\em Proc.Roy.Soc.
  London, Ser.A} {\bfseries 167} (1938) 148--169.

\bibitem[Frolov(1979)]{Frolov79}
V.~P. Frolov, ``The {N}ewman-{P}enrose method in the theory of general
  relativity'', in ``Problems in the General Theory of Relativity and the
  Theory of Group Representation'', N.~G. Basov, ed.
\newblock Consultants Bureau, 1979.

\bibitem[Geroch et~al.(1973)Geroch, Held, and Penrose]{Geroch73}
R.~Geroch, A.~Held, and R.~Penrose, ``A space-time calculus based on pairs of
  null directions'', {\em J. Math. Phys.} {\bfseries 14} (1973) 874--881.

\bibitem[Poisson(1999)]{Poisson:1999tv}
E.~Poisson, ``{An introduction to the Lorentz-Dirac equation}'',
 \href{http://xxx.lanl.gov/abs/gr-qc/9912045}{{\ttfamily gr-qc/9912045}}.

\bibitem[Teitelboim(1970)]{Teitelboim70}
C.~Teitelboim, ``Splitting of the maxwell tensor: Radiation reaction without
  advanced fields'', {\em Phys. Rev. D} {\bfseries 1} (1970), no.~6,
  1572--1582.

\bibitem[Bonnor(1974)]{Bonnor74}
W.~B. Bonnor, ``A new equation of motion for a radiating charged particle'',
  {\em Proc.R.Soc.Lond.A} {\bfseries 337} (1974) 591.

\bibitem[Landau and Lifshitz(1975)]{Landau75}
L.~Landau and E.~Lifshitz, ``The classical theory of fields'',
  Butterworth-Heinemann, Course of Theoretical Physics, Volume 2, {F}ourth
  revised english~ed., 1975.

\bibitem[Gralla et~al.(2009)Gralla, Harte, and Wald]{Gralla:2009md}
S.~E. Gralla, A.~I. Harte, and R.~M. Wald, ``{A Rigorous Derivation of
  Electromagnetic Self-force}'', {\em Phys.Rev.} {\bfseries D80} (2009) 024031,
   \href{http://xxx.lanl.gov/abs/0905.2391}{{\ttfamily arXiv:0905.2391}}.

\bibitem[Hille(1969)]{Hille69}
E.~Hille, ``Lectures on ordinary differential equations'', Addison-Wesley Pub.
  Co., 1969.

\bibitem[Hartman(1964)]{Hartman64}
P.~Hartman, ``Ordinary differential equations'', John Wiley \& Sons, Inc.,
  1964.

\end{thebibliography}

\begingroup\raggedright\endgroup

\end{document}